\documentclass[fleqn,usenatbib]{mnras}

\usepackage{newtxtext,newtxmath}

\usepackage[T1]{fontenc}

\DeclareRobustCommand{\VAN}[3]{#2}
\let\VANthebibliography\thebibliography
\def\thebibliography{\DeclareRobustCommand{\VAN}[3]{##3}\VANthebibliography}

\usepackage{graphicx}	
\usepackage{amsmath}	

\title[Peculiar-velocity cosmology with supernovae]{Peculiar-velocity cosmology with Types Ia and II supernovae}

\author[B. E. Stahl et al.]{Benjamin E. Stahl,$^{1,2}$\thanks{E-mail: benjamin\_stahl@berkeley.edu}\thanks{Marc J. Staley Graduate Fellow.}
Thomas de Jaeger,$^{3,1}$\thanks{Bengier Postdoctoral Fellow.}
Supranta S. Boruah,$^{4}$
\newauthor
WeiKang Zheng,$^{1}$
Alexei V. Filippenko,$^{1,5}$\thanks{Miller Senior Fellow.}
and Michael J. Hudson$^{6,7,8}$
\\
$^{1}$Department of Astronomy, University of California, Berkeley, CA 94720-3411, USA\\
$^{2}$Department of Physics, University of California, Berkeley, CA 94720-7300, USA\\
$^{3}$ Institute for Astronomy, University of Hawaii, 2680 Woodlawn Drive, Honolulu, HI 96822, USA\\
$^{4}$Steward Observatory, University of Arizona, Tucson, AZ 85719, USA\\
$^{5}$Miller Institute for Basic Research in Science, University of California, Berkeley, CA 94720, USA\\
$^{6}$Department of Physics and Astronomy, University of Waterloo, Waterloo, ON N2L 3G1, Canada\\
$^{7}$Waterloo Centre for Astrophysics, University of Waterloo, Waterloo, ON N2L 3G1, Canada\\
$^{8}$Perimeter Institute for Theoretical Physics, Waterloo, ON N2L 2Y5, Canada
}

\date{Accepted XXX. Received YYY; in original form ZZZ}

\pubyear{2021}

\begin{document}
\label{firstpage}
\pagerange{\pageref{firstpage}--\pageref{lastpage}}
\maketitle

\begin{abstract}
We present the Democratic Samples of Supernovae (DSS), a compilation of 775 low-redshift Type Ia and II supernovae (SNe~Ia~\&~II), of which 137 SN~Ia distances are derived via the newly developed snapshot distance method. Using the objects in the DSS as tracers of the peculiar-velocity field, we compare against the corresponding reconstruction from the 2M++ galaxy redshift survey. Our analysis --- which takes special care to properly weight each DSS subcatalogue and cross-calibrate the relative distance scales between them --- results in a measurement of the cosmological parameter combination $f\sigma_8 = 0.390_{-0.022}^{+0.022}$ as well as an external bulk flow velocity of $195_{-23}^{+22}$\,km\,s$^{-1}$ in the direction $(\ell, b) = (292_{-7}^{+7}, -6_{-4}^{+5})$\,deg, which originates from beyond the 2M++ reconstruction. Similarly, we find a bulk flow of $245_{-31}^{+32}$\,km\,s$^{-1}$ toward $(\ell, b) = (294_{-7}^{+7}, 3_{-5}^{+6})$\,deg on a scale of $\sim 30\,h^{-1}$\,Mpc if we ignore the reconstructed peculiar-velocity field altogether. Our constraint on $f\sigma_8$ --- the tightest derived from SNe to date (considering only statistical error bars), and the only one to utilise SNe~II --- is broadly consistent with other results from the literature. We intend for our data accumulation and treatment techniques to become the prototype for future studies that will exploit the unprecedented data volume from upcoming wide-field surveys.
\end{abstract}

\begin{keywords}
cosmology: observations, cosmological parameters -- supernovae: general -- methods: data analysis, statistical
\end{keywords}



\section{Introduction}
\label{sec:intro}

In observational cosmology, Type Ia supernovae (SNe~Ia) --- titanic explosions of white dwarfs in multistar systems \citep[e.g.,][]{Hoyle1960,Colgate1969,Nomoto1984} --- are highly prized for their standardiseability via the so-called ``width-luminosity relationship''\footnote{Here, ``width'' is with regard to an optical light curve, thus capturing the characteristic luminosity evolution timescale.} \citep[WLR; e.g.,][]{Phillips1993,Riess1996,mlcs2k2,SALT2,SNooPy}, which imbues them with the property of being precise extragalactic distance indicators. Exploiting this property with statistical samples of SNe~Ia has borne considerable fruit, including the discovery of the accelerating expansion of the Universe \citep{Riess1998,Perlmutter1999}, as well as the identification of a tension between local and distant measurements of the Hubble constant \citep[as reviewed by][]{RiessReview}. More recently, SNe~II --- colossal explosions of massive, evolved, hydrogen-envelope-bearing stars via core collapse \citep[see, e.g.,][for a review]{SmarttCC} --- have been used for such purposes \citep[e.g.,][]{SCM-TdJ,H0-TDJ} owing to their standardiseability via the standard candle method \citep[SCM;][]{hamuy02}.

In both sets of aforementioned cosmological analyses, recession velocities (i.e., redshifts) are compared to luminosity distances (i.e., distance moduli) to constrain the relevant cosmological parameters \citep[see, e.g.,][and references therein]{JhaReview}. There exists, however, a distinct set of cosmologically important parameters that can be probed not by the relationship between the aforementioned observables, but instead by the residuals that remain between them after removing a fiducial cosmological model \citep[e.g.,][]{Peebles,Huterer_cosmo,ScolnicWP}. Discounting unmodeled diversity in the underlying populations of SNe~Ia and SNe~II (which is still considerable in the latter), a significant contributor to these residuals comes from peculiar velocities --- i.e., deviations from the Hubble flow that are gravitationally induced by inhomogeneities. In turn, this makes SNe~Ia (and now, for the first time thanks to the SCM, SNe~II) an excellent probe of peculiar velocities \citep[e.g.,][]{MillerPV,Riess1997,WeyantPV,CF3}, as well as quantities derivable from them such as bulk flows in the nearby Universe \citep[which can shed light on the structures driving the flows of galaxies; e.g.,][]{RiessBF,ColinBF,DaiBF,FeindtBF,MathewsBF} and the cosmological parameter combination $f\sigma_8$ \citep[e.g.,][]{A1,HowlettLSSTSNIa,Huterer_SN,A2}. 

Though our analysis touches on each of these facets, our primary focus is on $f\sigma_8$, which can be decomposed as the dimensionless growth rate, $f = \frac{d\ln D}{d\ln a}$ \citep[where $D$ is the growth function of linear perturbations and $a$ is the scale factor; see][]{Peebles}, times the matter overdensity root-mean-squared fluctuations in a sphere of radius 8\,$h^{-1}$ Mpc, $\sigma_8$. The link back to peculiar velocities is provided by linear perturbation theory \citep{Peebles} via
\begin{equation}
    \mathbf{v}(\mathbf{r}) = \frac{H_0 f}{4\pi}\int d^3 \mathbf{r}^\prime \delta (\mathbf{r}^\prime)\frac{\mathbf{r}^\prime - \mathbf{r}}{|\mathbf{r}^\prime - \mathbf{r}|^3},
    \label{eq:perturbation}
\end{equation}
where $\delta = \rho/\overline{\rho} - 1$ is the overdensity field. Through $f \approx \Omega_m^\gamma$ \citep[where, e.g., $\gamma = 0.55$ for general relativity and a simple function of the dark energy equation of state in general; see][]{Linder07}, a constraint on $f\sigma_8$ can serve as a test of gravity \citep[e.g.,][]{Said_Gal}. As \citet{Linder13} show in their Figure 2, $f\sigma_8$ is most influenced by $\gamma$ at low redshifts. Thus, the use of SNe~II in our analysis (in addition to a large sample of SNe~Ia as described below) yields a tertiary benefit beyond (i) increasing statistical weight via a larger sample, and (ii) laying the foundation for future such analyses, of (iii) lowering the aggregate redshift of our full sample.

Still, as noted, the bulk of our sample is intentionally composed of SNe~Ia: they (with typical distance uncertainties of $\lesssim 10\%$ after standardisation via a WLR) offer more constraining power per unit object than do SNe~II \citep[with distance uncertainties of 10--15\% after applying the SCM;][]{SCM-TdJ} and galaxies \citep[with distance uncertainties of $\sim 25\%$ after applying scaling relationships; e.g.,][]{Said_Gal}. An unfortunate requirement, however, of applying a standard WLR-based standardisation to SN~Ia observations is that the requisite photometry must be sufficiently well sampled to reconstruct the ``width'' of the light curve. Consequently, there exists a significant set of SN~Ia observations that have not been used in cosmological analyses \citep[SNe~Ia are routinely cut for having too few epochs of photometry; e.g.,][]{Betoule2014,Foundation}.

With the advent of \texttt{deepSIP} \citep{deepSIP}, this requirement and the waste it incurs can be mitigated \emph{if} an optical spectrum is available --- by using a sophisticated convolutional neural network trained on a significant fraction of all relevant SN~Ia observations, \texttt{deepSIP} is able to map the spectrum of an SN~Ia to its corresponding light-curve shape with impressive precision. In turn, this has enabled the snapshot distance method \citep[SDM;][]{SDM}, which allows SN~Ia distances to be estimated with as little as one spectrum and two photometric points in different passbands. For the first time ever, we use the SDM to ``resurrect'' a significant sample of SNe~Ia which would otherwise have to be discarded in cosmological studies, and we include this sample in our analysis.

The remainder of this paper is organised as follows. First, we describe our accumulation of a sizeable dataset (Sec.~\ref{sec:data}) consisting of SNe~Ia and SNe~II, and then provide a comprehensive description of our analysis methodology (Sec.~\ref{sec:method}). We present our results in Section~\ref{sec:results} and then offer conclusions in Section~\ref{sec:conclusion}. Throughout we assume a flat $\Lambda$CDM cosmology with $\Omega_{m} = 0.3$ and $h = \mathrm{H}_0/(100\,\mathrm{km\,s^{-1}})$, where H$_0$ is the local Hubble constant.

\section{Data}
\label{sec:data}

\subsection{Type Ia Supernovae}

\subsubsection{Amended Second Amendment SN~Ia Compilation}

As a starting point we turn to the --- until now --- largest ever peculiar-velocity catalogue derived from nearby SNe~Ia: the Second Amendment (A2) compilation of 465 SN~Ia distances \citep{A2}. A2 draws SNe~Ia from the third Carnegie Supernova Program (CSP) data release \citep{CSP3}, the Lick Observatory Supernova Search (LOSS) cosmology sample \citep{G12}, the first Foundation Supernova Survey data release \citep{Foundation}, and the First Amendment (A1) compilation \citep{A1}, which is itself an aggregation of SNe~Ia from the first CSP data release \citep{CSP1a} and the ``Constitution'' set  \citep{Hicken2009}, atop which the ``Amendments'' are made.

Though nicely organised and homogenised in many regards, the A2 compilation remains fundamentally heterogeneous. For example, the A1 and CSP DR3 subcatalogs consist of distance moduli (derived using \emph{different} light-curve fitters), while the Foundation and LOSS subcatalogs provide only SALT2 \citep{SALT2} light-curve parameters\footnote{The parameters derived from fitting the SALT2 model to a multipassband temporal series of SN~Ia fluxes (i.e., a multiband light curve) are $m_B$, the observed $B$-band magnitude at maximum light; $x_1$, the light-curve stretch parameter; and $c$, the colour parameter.}, from which distances are derived using the \citet{Tripp} formula,
\begin{equation}
    \mu = m_B - M + \alpha x_1 - \beta_\textrm{SN} c,
    \label{eq:tripp}
\end{equation}
where we have added the ``SN'' subscript to $\beta$ to avoid any confusion with $\beta = f/b$ used elsewhere in our analysis. Our analysis method (see Sec.~\ref{sec:method}) is therefore carefully formulated to account for these and other differences between subcatalogs, ensuring (among other things) a consistent relative distance scale and proper weighting.

For the analysis described herein we use an \emph{amended} version of A2 (hereafter A2.1), derived by removing\footnote{In cases of duplicated SNe~Ia, we choose which set of data to keep by using the following order of preference (highest first): CSP DR3, CSP DR1 (in A1), LOSS, Constitution (in A1).} all duplicates of individual SNe~Ia from A2. Though scientifically important in the sense that it prevents any particular peculiar-velocity signal from having an artificially high weight in our fit, this removes just 13/465 distances. As a result, prior and future analyses that use the full A2 set (instead of the 452 in A2.1) should only suffer from a minimal (perhaps imperceptible) bias due to duplication.

\subsubsection{Lick Observatory Supernova Search Sample}

In the time since the publication of the first LOSS data release \citep{Ganeshalingam2010} --- the source for many of the low-redshift objects used in the LOSS cosmology sample --- the LOSS team has continued to perform multiband photometric and spectroscopic follow-up observations of newly discovered SNe~Ia \citep{S19,S20,deepSIP}. We draw from these new data to construct LOSS2.0: a set of 45 distinct (from A2.1) SNe~Ia that pass minimal cuts for utility and reliability. In contrast to the SALT2 parameters of LOSS1.0 (from A2.1), LOSS2.0 is a catalog of distance moduli taken directly from the \texttt{SNooPy} \citep{SNooPy} fits that were published in accompaniment to the light curves from which they were derived, and then uniformly shifted to be consistent with our fiducial distance scale. The corresponding redshifts are all firmly established from SN host galaxies and in the CMB frame, except that those objects known to be within a group of galaxies have their redshifts updated to that of their host group. This is consistent with the preparation of A2 \citep[and hence A2.1;][]{A2}, and helps to mitigate the noise induced by the velocity dispersion between member galaxies.

\subsubsection{Snapshot Distances Sample}
\label{sssec:sdm}

The amalgamation described above (A2.1 $+$ LOSS2.0) is, perhaps, the largest ever compilation of low-redshift SN~Ia distances ever assembled, utilising the collective observations of many campaigns over the last $30+$ years. With conventional methods (i.e., applying a WLR to well-sampled SN~Ia light curves), one would find that this set already accounts for the majority of data that are publicly available. However, this set \emph{can} be further expanded by employing the SDM, which removes the requirement that light curves be well sampled in exchange for providing an optical spectrum. We therefore devote this section to describing the accumulation of a significant sample of SDM-derived distances for sparsely observed SNe~Ia.

To do so, we turn to the Open Supernova Catalog\footnote{\url{https://sne.space}} \citep[OSC;][]{openSNe} which, as of the date of our original query, contained a total of 64,715 transients with 14,140 classified as some form of SN~Ia. From this subset, we impose three additional cuts to derive our initial candidate pool for the SDM: (i) we keep only those 6359 that have redshift $z < 0.07$ as required by our analysis method (see Sec.~\ref{sec:method}), (ii) of these, we keep those 3951 that have at least one spectrum, and (iii) we keep those 2900 objects that, in addition to the above, have at least two epochs of photometry. We emphasise that these cuts yield only an \emph{initial} sample of candidates that satisfy the most basic prerequisites of our analysis.

Following a sequence of judiciously chosen cuts to ensure quality, consistency, and compatibility (see Appendix~\ref{app:sdm} and Figure~\ref{fig:cuts}), we perform the SDM and as a check, a na\"ive fit to the light curves using no \texttt{deepSIP}-derived information. The top-line results are consistent with our high expectations for the power of the SDM: out of the 223 objects considered, the SDM succeeds in deriving a distance in 197 cases while the standard fit succeeds in only 175 (owing mostly to data sparsity). Moreover, we find a high degree of consistency between SDM distances and their conventionally derived counterparts when both methods are successful --- a Kolmorgorov-Smirnoff test produces a $p$-value of 0.993, and the distances have a median residual of just 0.001\,mag.

\begin{figure}
    \includegraphics[width=\columnwidth]{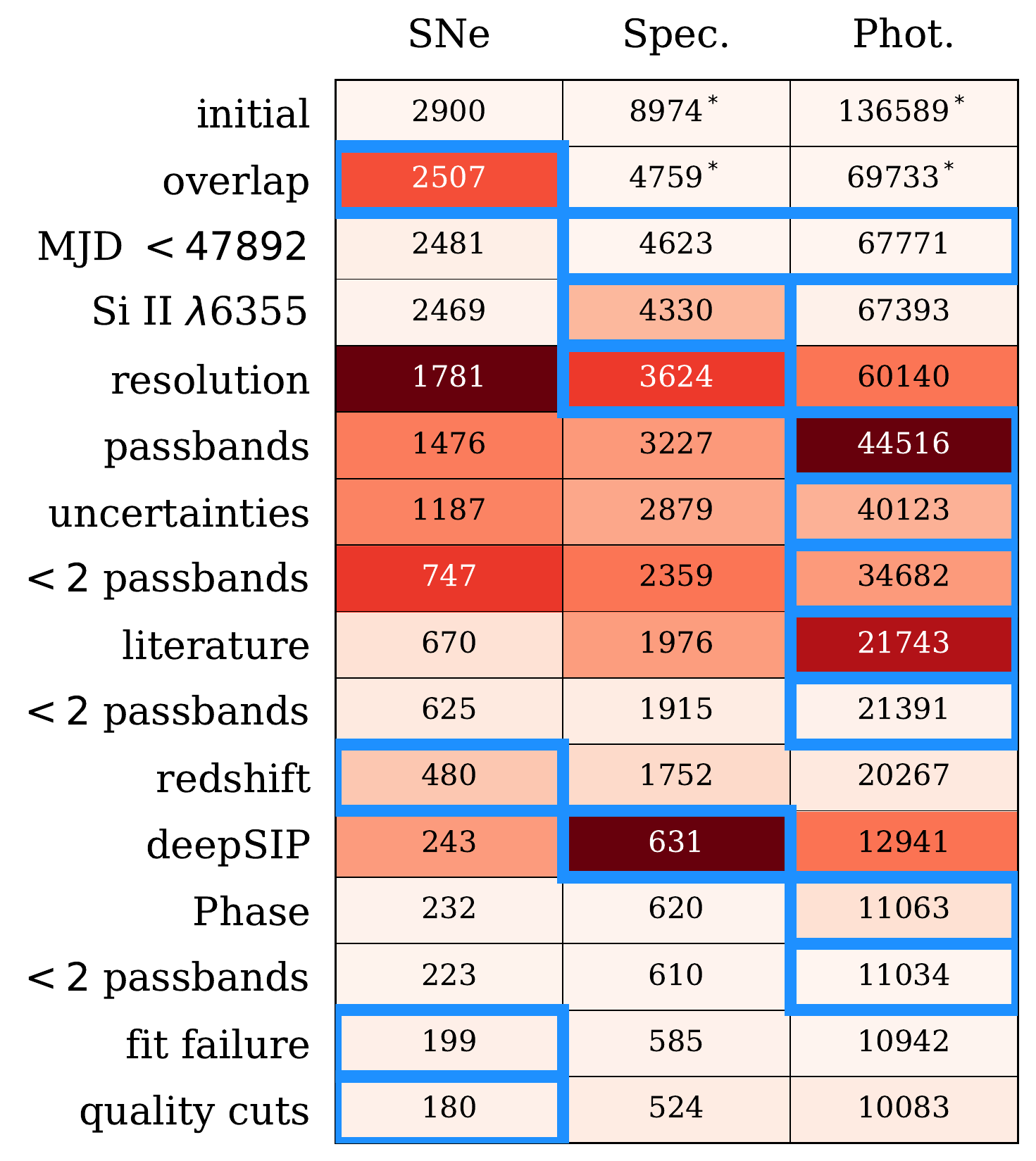}
    \caption{Full accounting of the cuts that take the initial sample of Open-Supernova-Catalog-obtained SN~Ia candidates for snapshot distances to the final sample of robust, high-confidence distances. The colour map shows the size of each cut (per column) as a percentage of the number initially retrieved. Numbers marked with an asterisk (*) are omitted from the calculations because they are self-reported by the Open Supernova Catalog. Boxes are used to denote the data level (e.g., SNe, spectra, photometry) to which a given cut is applied. Repeated cuts are intentional and explained in Appendix~\ref{app:sdm}.}
    \label{fig:cuts}
\end{figure}

Still, there are several instances where the difference between an individual SDM distance and its corresponding conventional estimate do differ significantly. This is not unexpected, given the highly underconstrained nature of the conventional fits in the presence of sparse data; however, to be cautious, we visually inspect the fits for all SNe whose distance estimates (SDM and conventional) differ by more than the lesser of their corresponding error bars. Of the 34 SNe for which this is the case, we manually assign the distance used in our analysis to be from the SDM in 19 cases, from the conventional fit in one case, and for the remainder, to be selected using our default strategy of using the distance with a smaller uncertainty. All told, this distills our OSC sample into 199 SNe, 150 of which have distances derived using the SDM (the remainder are derived via the conventional method).

After uniformly shifting all distances to our fiducial distance scale, we impose minimal reliability cuts that reduce the sample to 193 objects, and we cut a further 13 objects for having distance moduli that differ by more than 1\,mag from the cosmologically-expected value at their redshifts. In our sample, this preferentially removes fainter-than-expected objects --- an encouraging result in light of the fact that our selection criteria do not include a spectroscopic classification step to explicitly identify and remove contaminating transient events (such as core-collapse SNe) that may be spuriously present. This results in a final sample of 180 distances (137 of which are derived from the SDM).

\subsection{Type II Supernovae}

With the addition of the OSC sample, our low-redshift SN~Ia distance compilation provides \emph{comprehensive} coverage of all publicly available objects that satisfy basic suitability criteria (note that these criteria are now agnostic to how well sampled an individual object's light curve is, thanks to the SDM). Without proprietary datasets, one would be unable to meaningfully increase the sample size at the present time. With this limitation in mind and a persisting desire to grow the sample larger while simultaneously building a foundation for further work, we turn to another class of standardiseable SNe.

In contrast to SNe~Ia (which are standardised via a WLR), SNe~II can be standardised using the SCM \citep{hamuy02,SCM-2017}, which exploits the empirical fact that intrinsically brighter SNe~II have higher expansion velocities (as probed by the H$\beta$ spectral feature velocity, $v{_\mathrm{H\beta}}$) and are bluer in colour ($c$). As such, one can write (similar to Eq.~\ref{eq:tripp})
\begin{equation}
    \mu = m - M + \alpha \log_{10}\left(\frac{v_{\mathrm{H}\beta}}{\overline{v}_{\mathrm{H}\beta}}\right) - \beta_\textrm{SN} (c - \overline{c}),
    \label{eq:tripp-SNeII}
\end{equation}
where $m$ is the apparent magnitude in a given passband at 43\,d after the explosion and the overbars are used to denote averaged quantities. As in Equation~\ref{eq:tripp}, the absolute magnitude ($M$) and slopes ($\alpha$ and $\beta_\textrm{SN}$) are nuisance parameters to be derived jointly with the scientifically important parameters in our analysis (see Sec.~\ref{sec:method} for more details).

We therefore compile a sample of 98 low-redshift SNe~II --- the first ever to be used in a peculiar-velocity analysis --- from the following surveys: CSP-I \citep[][]{ham06}, LOSS \citep{dejaeger19}, the Sloan Digital Sky Survey-II SN Survey \citep[SDSS-II;][]{SDSSSNe,andrea10}, and the Dark Energy Survey Supernova Program \citep[DES-SN;][]{bernstein12,SCM-TdJ}. More information about our SN~II compilation is provided by \citet{H0-TDJ}. Consistent with the other samples used herein, we use group redshifts for any SNe~II known to be in a group of galaxies.


\subsection{Final Compilation: The Democratic Samples of SNe}

\begin{figure}
    \includegraphics[width=\columnwidth]{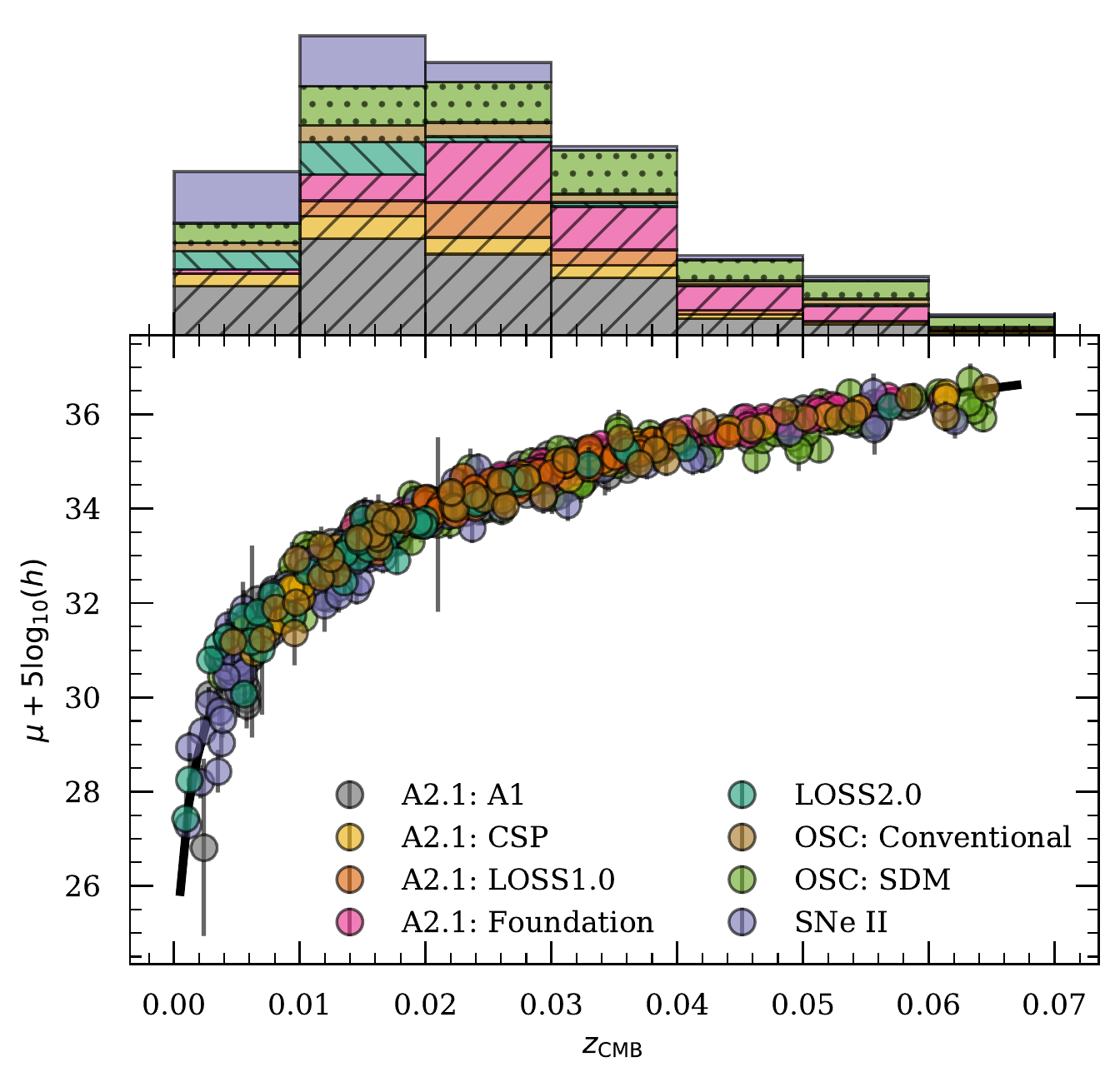}
    \caption{Hubble diagram with redshift distribution projected above for our DSS sample, with subcatalogs distinguished by colour. Distance moduli and error bars (which include subcatalog-specific instrinsic scatter) are derived following the prescription of Section~\ref{sec:method}.}
    \label{fig:zhist}
\end{figure}

Aggregating over all sources listed above (A2.1 + LOSS2.0 + OSC + SNe~II), we arrive at a catalog of 775 SN-based distances. As it is built atop the ``Constitution'' set and its ``first'' and ``second'' amendments, we dub this catalog the ``Democratic Samples of Supernovae'' (DSS). The DSS represents a $\sim 70\%$ increase over the number of unique objects in A2 (i.e, the number of SNe~Ia in A2.1), which was, until now, the largest such catalog used to study bulk flows in the nearby Universe. Moreover, our sample is the first \emph{ever} to use SNe~II for this purpose. In addition to opening an entirely new avenue with which to study peculiar velocities, our SN~II subcatalog benefits our analysis by lowering the aggregate redshift (as is clearly visible in Fig.~\ref{fig:zhist}) and  characteristic depth, 
\begin{equation}
    d_* = \frac{\sum_i r_i/\sigma_i^2}{\sum_i 1/\sigma_i^2} ,
\end{equation}
of our full sample (e.g., $d_*^\mathrm{A2.1} = 39\,h^{-1}$\,Mpc, $d_*^\mathrm{SNe\,II} = 10\,h^{-1}$\,Mpc, and $d_*^\mathrm{DSS} = 28\,h^{-1}$\,Mpc). As noted in Section~\ref{sec:intro}, our analysis has the most discriminating power between gravitational models at low redshifts.

\begin{figure}
    \includegraphics[width=\columnwidth]{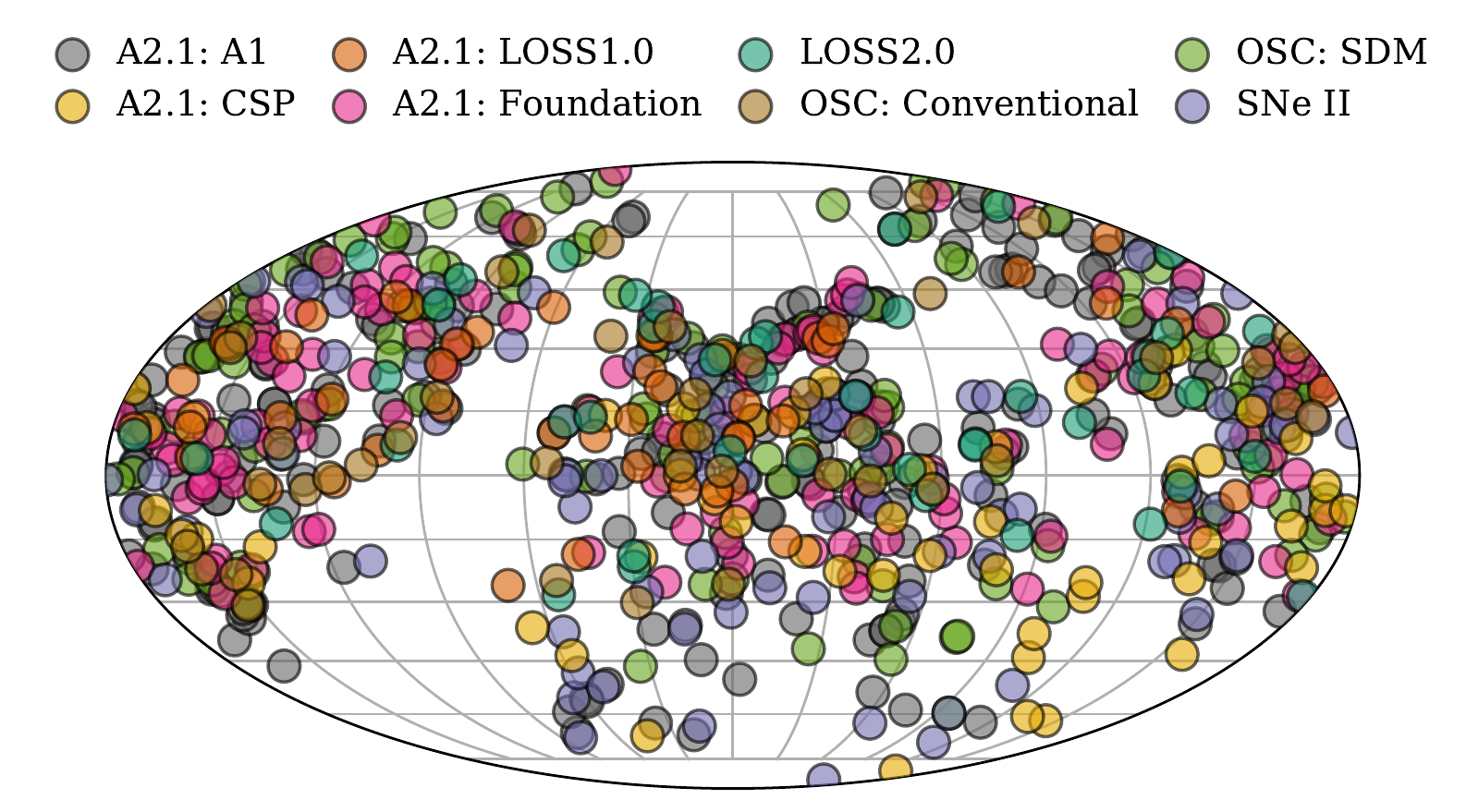}
    \caption{Mollweide projection of the on-the-sky distribution of our sample (in equatorial coordinates), with subcatalogs distinguished by colour. Unlike some prior studies, our Southern-hemisphere coverage is substantial.}
    \label{fig:sky}
\end{figure}

A secondary benefit of including the SN~II subcatalog in our analysis is the added Southern-hemisphere coverage that it yields (see Fig.~\ref{fig:sky}, though our SN~Ia sample also fares much better than prior samples in this regard). With upcoming wide-field surveys, we expect sky coverage to become even better in short order, and  especially for studies that use SNe~Ia (both well and sparsely observed) and SNe~II as we do here.

\subsection{Reconstructed Density and Velocity Fields}
\label{ssec:reconstruction}

As will be shortly understood, our analysis methodology requires \emph{observed} peculiar velocities (e.g., our DSS catalog) and knowledge of the overdensity field ($\delta$) which, in light of Equation~\ref{eq:perturbation}, dictates the peculiar-velocity field for a given set of cosmological parameters. There are, however, two problems that prevent the direct application of Equation~\ref{eq:perturbation} in our (and indeed, every) case:
\begin{enumerate}
    \item $\delta$ is not observable. The density contrast as traced by galaxies ($\delta_g$) is, however, and we assume the two are related by $\delta_g = b\delta$, where $b$ is the linear bias factor.
    \item $\delta_g$ cannot be measured for all space. Instead, we assume it is measured only up to some maximum distance, $R_\mathrm{max}$, by (for example) an all-sky redshift survey.
\end{enumerate}
Invoking the above and defining $\beta \equiv f/b$, we modify Equation~\ref{eq:perturbation} to
\begin{equation}
    \mathbf{v}(\mathbf{r}) = \frac{H_0 \beta}{4\pi}\int_0^{R_\mathrm{max}} d^3 \mathbf{r}^\prime \delta_g (\mathbf{r}^\prime)\frac{\mathbf{r}^\prime - \mathbf{r}}{|\mathbf{r}^\prime - \mathbf{r}|^3} + \mathbf{V}_\mathrm{ext},
    \label{eq:modified_perturbation}
\end{equation}
where the first term is the peculiar-velocity field generated by structure within the volume covered by the redshift survey, and the second term ($\mathbf{V}_\mathrm{ext}$) is a coherent, residual bulk flow driven by structure \emph{outside} the covered volume. Together, then, these terms constitute a \emph{reconstructed} peculiar field that can be compared with direct measurements (as we do here) to constrain $\beta$ and $\mathbf{V}_\mathrm{ext}$. This method has been validated with $N$-body simulations and semi-analytic galaxy formation models; bias in the resulting values of $f \sigma_8$ (as derived from $\beta$) are found to be $\lesssim 5$\% \citep{HollingerHudsonVV}. We note that all uncertainties reported herein are statistical only.

As a matter of implementation, we derive $\delta$ and $\mathbf{v}$ (both as a function of position) by interpolating over the reconstructions\footnote{The reconstructions are available at \url{https://cosmicflows.iap.fr}.} produced by \citet{Carrick} up to $R_\mathrm{max} = 200$\,$h^{-1}$\,Mpc using the 2M++ galaxy redshift catalog \citep{2mpp}. For $\mathbf{v}$, we remove the coherent bulk flow, $\mathbf{V}_\textrm{ext} = (89, -131, 17)$\,km\,s$^{-1}$ in galactic Cartesian coordinates, and normalise over $\beta = 0.43$, included in their published velocity field, because our aim is to refit those quantities here. Hereafter, $\mathbf{v}$ will refer to the peculiar-velocity reconstruction with this modification.

\section{Method}
\label{sec:method}

The fundamental aim of this study is to compare the reconstructed peculiar-velocity field (see Sec.~\ref{ssec:reconstruction}) to the tracers which comprise our DSS catalog, and in doing so, to fit for $\beta$ and  $\mathbf{V}_\textrm{ext}$. The former allows us to constrain the degenerate cosmological parameter combination $f\sigma_8$ (see Sec.~\ref{sec:results}), while the latter may result from, and thus constrain, properties of large-scale structures that exist beyond the volume of our selected peculiar velocity field reconstruction. We devote the remainder of this section to describing our method for fitting these quantities, which we denote collectively as the \emph{flow model}, as well as a number of subcatalog-specific nuisance parameters that homogenise the objects in our global catalog.

\subsection{The Forward Likelihood Method}
\label{ssec:fwd_lkl}

We use an updated implementation of the \emph{forward likelihood method} \citep{fwd_lkl} used by \citet{A2}. In this method, the flow model ($\beta, \mathbf{V}_\textrm{ext}$) and (mostly subcatalog-specific) nuisance parameters ($\mathbf{\Theta}$) are coupled to the observable parameters of an SN ($\mathbf{x}_i$) through the conditional probability $\mathcal{P}(\beta, \mathbf{V}_\textrm{ext}, \mathbf{\Theta} | \mathbf{x}_i)$, which, given Bayes' theorem, can be expressed as
\begin{equation}
    \mathcal{P}(\beta, \mathbf{V}_\textrm{ext}, \mathbf{\Theta} | \mathbf{x}_i) \propto \mathcal{P}(\mathbf{x}_i | \beta, \mathbf{V}_\textrm{ext}, \mathbf{\Theta}) \mathcal{P}( \beta, \mathbf{V}_\textrm{ext}, \mathbf{\Theta}) . \label{eq:bayes}
\end{equation}
An advantage of this method is that it mitigates inhomogeneous Malmquist bias, which arises when line-of-sight inhomogeneities are ignored \citep{1995PhR...261..271S} and can cause the inferred value of $\beta$ to be biased high \citep[e.g.,][]{Carrick}. This is accomplished by accounting for inhomogeneities along the line of sight via the radial distribution
\begin{equation}
\mathcal{P}(\mathbf{r}|\mathbf{\Theta}) = \frac{1}{\mathcal{N}(\mathbf{\Theta})}r^2 \exp\left\{-\frac{\left[\mu(r) - \mu(\mathbf{\Theta})\right]^2}{2\sigma^2_\mu(\mathbf{\Theta})}\right\}\left[1 + \delta_g(\mathbf{r})\right],
    \label{eq:radial}
\end{equation}
where $\mathbf{r}$ is a vector whose length ($r$) denotes a comoving distance and whose direction corresponds to a point on the celestial sphere, $\mathcal{N}(\mathbf{\Theta})$ is a nuisance-parameter-dependent normalisation factor, $\delta_g$ is the overdensity in the galaxy field, and $\sigma_\mu$ --- the quadrature-sum of the (fitted, subcatalog-specific) intrinsic scatter and the propagated distance uncertainty --- is explicitly shown as a function of $\mathbf{\Theta}$ for clarity. This is then marginalised over to derive the likelihood
\begin{equation}
    \mathcal{P}(\mathbf{x}_i | \beta, \mathbf{V}_\textrm{ext}, \mathbf{\Theta}) = \int_0^{R_\mathrm{max}} dr \mathcal{P}(\mathbf{x}_i | \mathbf{r}, \beta, \mathbf{V}_\textrm{ext}, \mathbf{\Theta}) \mathcal{P}(\mathbf{r}|\mathbf{\Theta}) ,
    \label{eq:marginalization}
\end{equation}
where $R_\mathrm{max}$ corresponds to the extent of our reconstructed density and peculiar velocity fields, and
\begin{equation}
    \mathcal{P}(\mathbf{x}_i | \mathbf{r}, \beta, \mathbf{V}_\textrm{ext}, \mathbf{\Theta}) = \frac{1}{\sqrt{2\pi \sigma_v^2}}\exp\left\{-\frac{\left[cz_\mathrm{obs} - cz_\mathrm{pred}(\mathbf{r}, \beta, \mathbf{V}_\textrm{ext}) \right]^2}{2\sigma_v^2}\right\}.
\end{equation}
In our analysis, we treat $\sigma_v$ as a global nuisance parameter (i.e., a component of $\mathbf{\Theta}$) to be fit simultaneously with the other parameters of our model \citep[in contrast to other studies that fix $\sigma_v$ to (for example) 200\,km\,s$^{-1}$ or 150\,km\,s$^{-1}$;][respectively]{fwd_lkl,A2}, take $z_\mathrm{obs}$ as a mandatory component of $\mathbf{x}_i$ (prepared as discussed in Sec.~\ref{sec:data}), and compute the predicted redshift (shown above with its explicit parameter dependence) according to \citep{DavisRedshiftPV}
\begin{equation}
    1 + z_\mathrm{pred} = \left[1 + \tilde{z}(r)\right]\left\{1 + \frac{1}{c}\left[\beta \mathbf{v}(\mathbf{r}) + \mathbf{V}_\mathrm{ext}\right]\cdot \hat{\mathbf{r}}\right\},
    \label{eq:zpred}
\end{equation}
where $\mathbf{v}$ is the reconstructed peculiar velocity and $\tilde{z}$ is the cosmological redshift, approximated to second order \citep{Peebles} as
\begin{equation}
    \tilde{z} = \frac{1}{1 + q_0}\left[1 - \sqrt{1 - \frac{2H_0 r}{c}(1 + q_0)}\right],
\end{equation}
which makes it clear that $\tilde{z}$ is solely a function of the assumed cosmological model and a precise, redshift-independent SN distance (parameterised here by the comoving distance, $r$). For the deceleration parameter, we use $q_0 = \Omega_m/2 - \Omega_\Lambda$ for a flat $\Lambda$CDM Universe.

\subsection{Observable Parameters, Nuisance Parameters, and Priors}
\label{ssec:params}

Thus far, we have maintained a level of abstraction from the data-specific details of our method, but in order to continue our development, we must now delve into them. Namely, we have referred to the observable parameters of a given SN only as $\mathbf{x}_i$ and nuisance parameters (with the exception of $\sigma_v$) as $\mathbf{\Theta}$. Moreover, we have not yet addressed the prior probability term in Equation~\ref{eq:bayes}, $\mathcal{P}( \beta, \mathbf{V}_\textrm{ext}, \mathbf{\Theta})$. There are three specific cases to describe, but first we delineate those attributes that are generically present, regardless of the case.

As was stated in Section~\ref{ssec:fwd_lkl}, $z_\mathrm{obs}$ is a mandatory component of all $\mathbf{x}_i$. This is also true for object coordinates (right ascension and declination), from which $\hat{\mathbf{r}}$ (i.e., the direction of $\mathbf{r}$) is computed. The remaining components of a given $\mathbf{x}_i$ --- which are responsible for providing a redshift-independent distance --- depend on the type of subcatalog to which it belongs. We delve into this, along with nuisance parameters and priors, in the following paragraphs.

\subsubsection{``Simple Distance'' Subcatalogs}

``Simple distance'' subcatalogs are those which directly include distance moduli (e.g., A1, CSP, LOSS2.0, OSC). These distance moduli (and their corresponding error bars) are converted into comoving distances (with propagated error bars), which then become the final components of the vector of observables for a given SN, $\mathbf{x}_i$.

For such catalogs there are two nuisance parameters, $\mathbf{\Theta} = (\eta, \sigma_\mathrm{int})$, that are jointly fit with the flow model. The first, $\eta$\footnote{Our $\eta$ plays the same role as the $\tilde{h}$ used by \citet{A2}.}, rescales the reported distance as $r\to \eta r$, ensuring a globally consistent relative distance scale across multiple subcatalogs. The second, $\sigma_\mathrm{int}$, is the usual (fitted) intrinsic scatter employed in SN analyses to account for unmodeled behaviour that is not captured in the standardisation technique. In our analysis, $\sigma_\mathrm{int}$ serves a secondary purpose of controlling the weight of a given subcatalogue in the global fit (because each catalogue has its own fitted $\sigma_\mathrm{int}$).

For $\eta$, we impose a simple, positive-only prior and for $\sigma_\mathrm{int}^2$ we use a log-normal prior with a peak corresponding to $\sigma_\mathrm{int} = 0.15$\,mag. The latter enforces the requirement that $\sigma_{\mathrm{int}}$ be positive while simultaneously being flexible enough to accommodate the range of values expected for both conventionally- and SDM-standardised SNe~Ia.

\subsubsection{SN~Ia Tripp Distance Subcatalogues}

For the remaining SN~Ia subcatalogues in the DSS (e.g., LOSS, Foundation), we add the fitted SALT2 parameters $(m_B, x_1, c)$ and their uncertainties as the final components of the vector of observable quantities for a given SN, $\mathbf{x}_i$. Distance moduli are then derived via Equation~\ref{eq:tripp} with the requisite nuisance parameters being jointly fit with the flow model and drawn from $\mathbf{\Theta} = (M, \alpha, \beta_\mathrm{SN}, \sigma_\mathrm{int})$. Using concepts from differential calculus, we express the propagated distance modulus error (including the intrinsic scatter term) as
\begin{equation}
    \sigma_\mu^2 = \sigma_{m_B}^2 + (\alpha \sigma_{x_1})^2 + (\beta_\mathrm{SN} \sigma_c)^2 + \sigma_\mathrm{int}^2.
    \label{eq:tripp-error}
\end{equation}
Finally, we convert the distance moduli (and their corresponding error bars) into comoving distances (with propagated error bars). From a practical point of view, the peak \emph{absolute} magnitude, $M$, serves a role analogous to $\eta$, in the sense that its value brings the relative distance scale of the subcatalogue for which it is fit into agreement with the global scale used in our analysis.

We place only very simple and nonrestrictive priors on $M, \alpha,$ and $\beta_\mathrm{SN}$, requiring $M < 0$ and $\alpha, \beta_\mathrm{SN} > 0$. For $\sigma_\mathrm{int}$ we impose the same log-normal prior as for simple distance subcatalogues. 

\subsubsection{SN~II Subcatalogue}

Parallel to our treatment of SN~Ia Tripp Distance Subcatalogues, we add the parameters (and their uncertainties) required by Equation~\ref{eq:tripp-SNeII} (i.e., $m, v_{\mathrm{H}\beta}, c$) for deriving SN~II distance moduli as the final components  of the vector of observable quantities for a given object, $\mathbf{x}_i$. Propagating all uncertainties and including the intrinsic scatter term, the distance modulus error can be expressed (in light of Eq.~\ref{eq:tripp-SNeII}) as
\begin{equation}
    \sigma_\mu^2 = \sigma_{m}^2 + \left(\frac{\alpha \sigma_{v_{\mathrm{H}\beta}}}{\ln(10) v_{\mathrm{H}\beta}}\right)^2 + (\beta_\mathrm{SN} \sigma_c)^2 + \sigma_\mathrm{int}^2.
    \label{eq:tripp-SNII-error}
\end{equation}
As a result, the nuisance parameters for our SN~II Tripp Distance Subcatalogue are $\mathbf{\Theta} = (M, \alpha, \beta_\mathrm{SN}, \sigma_\mathrm{int})$, consistent with the previous case. We therefore impose the same priors, except that the peak of the $\sigma_\mathrm{int}$ prior is shifted to 0.27\,mag, consistent with the result found by \citet{H0-TDJ}.

\subsection{Implementation}

Given the definitions above and the dataset described in Section~\ref{sec:data}, we are now able to implement our model and derive the best-fitting flow model (and nuisance parameters). Formally, these optimal values are, in light of Equation~\ref{eq:bayes}, the ones that given our entire DSS catalogue, $\{\mathbf{x}\}$, maximise the joint posterior (assuming statistical independence),
\begin{equation}
    \mathcal{P}(\beta, \mathbf{V}_\textrm{ext}, \mathbf{\Theta} | \{\mathbf{x}\}) \propto \mathcal{P}(\beta, \mathbf{V}_\textrm{ext}, \mathbf{\Theta})\prod_i \mathcal{P}(\mathbf{x}_i | \beta, \mathbf{V}_\textrm{ext}, \mathbf{\Theta}) . \label{eq:likelihood}
\end{equation}
In practice, it is more convenient to work with the logarithm of this which we define up to a multiplicative constant as $\mathcal{L}$, so that (using logarithm algebra)
\begin{equation}
    \mathcal{L} = \ln \mathcal{P}( \beta, \mathbf{V}_\textrm{ext}, \mathbf{\Theta}) +  \sum_i \ln\mathcal{P}(\mathbf{x}_i | \beta, \mathbf{V}_\textrm{ext}, \mathbf{\Theta}) .
    \label{eq:loglikelihood}
\end{equation}
The flow model and nuisance parameters are thus inferred by sampling \citep[using the \texttt{emcee} package;][]{emcee} assuming uniform priors on $\beta > 0$ and $\mathbf{V}_\mathrm{ext}$, and a reasonably broad ($\sigma = 15$\,km\,s$^{-1}$) Gaussian prior on $\sigma_v$ centred at $150$\,km\,s$^{-1}$ \citep[the static value used by other studies; e.g.,][]{Carrick,A2}. In this Markov Chain Monte Carlo (MCMC) simulation, we find that 256 walkers and 2000 steps (after removing 500 for ``burn in'') yields robust convergences of all parameters. We describe the results in the following section.

\section{Results}
\label{sec:results}

As discussed in Section~\ref{sec:method}, our method of analysis results in best-fit sets of parameters that broadly belong to two categories: (i) subcatalogue-specific nuisance parameters, and (ii) the flow model, which itself bears cosmological significance. Prior to discussing the latter and its implications, we investigate the former as derived from our DSS catalogue, and through this validation exercise, we strengthen the weight of all subsequent conclusions.

\subsection{Nuisance Parameters}

\begin{figure*}
    \includegraphics[width=\textwidth]{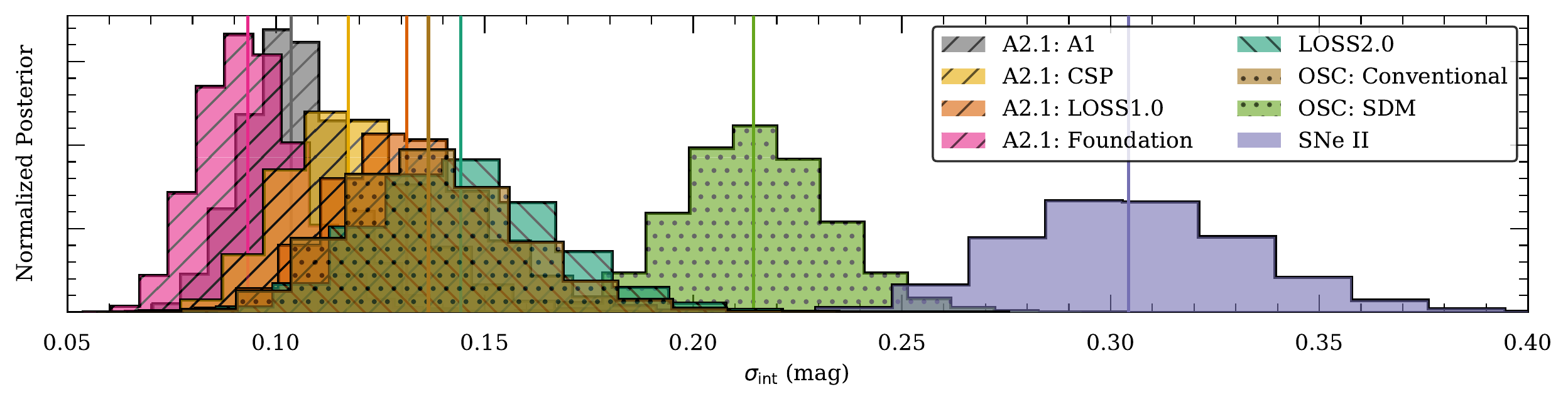}
    \caption{Posterior distributions for each subcatalogue's $\sigma_\mathrm{int}$ term. Each is normalised to unit area and as a result, distributions with lower peaks are necessarily broader. The distinct populations are visually apparent: (in order of increasing $\sigma_\mathrm{int}$) those subcatalogues with WLR-determined SN~Ia distances, SDM-derived SN~Ia distances, and SCM-derived SN~II distances. Colour-coded vertical lines are used to denote the median of each posterior.}
    \label{fig:sigma}
\end{figure*}

The most straightforward of all the subcatalogue-specific nuisance parameters to compare is the intrinsic scatter term, $\sigma_\mathrm{int}$, because it is present in each set. We perform such a comparison by visualising the posterior distribution for each subcatalogue's $\sigma_\mathrm{int}$ term (as derived from our MCMC analysis) in Figure~\ref{fig:sigma}. As expected, we find that our SN~II subcatalogue exhibits the largest scatter at $0.30_{-0.03}^{+0.03}$\,mag (median value with 16th and 84th percentile differences; reported here and throughout), consistent\footnote{Here and henceforth, we consider two measurements to be consistent if the lesser plus its upper uncertainty bound exceeds the greater minus its lower uncertainty bound.} with the corresponding determination made with a superset of the subcatalogue in a different application \citep[$0.27_{-0.04}^{+0.04}$\,mag;][]{H0-TDJ}.

Turning to those SN~Ia subcatalogues with distances derived via conventional WLRs (i.e., all but our SDM subcatalogue), we note that all have $\sigma_\mathrm{int}$ posteriors that peak within the vicinity of $\sim 0.1$\,mag, consitent with recent analyses that have used related datasets \citep[e.g.,][]{Foundation,BurnsH0,A2}. Finally, we note that our subcatalogue of SDM-derived SN~Ia distances has a $\sigma_\mathrm{int}$ posterior that falls nicely within a SN~Ia -- SN~II continuum, peaking at $0.21_{-0.02}^{+0.02}$\,mag. This represents a further $\sim 0.16$\,mag in (quadrature-added) scatter relative to the conventional WLR OSC catalogue and lies within the range found by \citet{SDM} when validating the SDM.

To compare other nuisance parameters across distinct subcatalogues is less straightforward given the differences in parameters for different subcatalogue types (e.g., $\eta$ for ``Simple Distance'' subcatalogues, and $M, \alpha, \beta_\mathrm{SN}$ for SN~Ia and SN~II Tripp Distance subcatalogues). Still, we find sensible results which we summarise by way of the following comments.
\begin{enumerate}
    \item The parameters of our SN~Ia Tripp Distance subcatalogues are all consistent (within the uncertainties) with the corresponding values found by \citet{A2}.
    \item Our SN~II Distance subcatalogue's $\alpha$ and $\beta_\mathrm{SN}$ values agree (within the uncertainties, albeit just barely for the former) with those found by \citet{H0-TDJ}, but $M$ does not. However, as Figure 3 from \citet{H0-TDJ} reveals a strong positive correlation between H$_0$ and $M$, a weaker (but clearly present) negative correlation with $\alpha$, and negligible correlation with $\beta_\mathrm{SN}$, our lack of consistency with $M$ is not a concern; rather, it is an expected consequence of our use of a different relative distance scale.
    \item Our fitted values of $\eta$ for the three Simple Distance subcatalogues that have distances derived via the same light-curve fitter (i.e., LOSS2.0 and both OSC subcatalogues) are consistent with one another, given their uncertainties.
\end{enumerate}
The only global nuisance parameter in our analysis is $\sigma_v$, for which we find $\sigma_v = 128_{-9}^{+10}$\,km\,s$^{-1}$ --- lower than 150\,km\,s$^{-1}$, but not significantly so given the scale of the prior placed on it. Regardless, the flow-model results are not particularly sensitive to the exact value owing to a degree of degeneracy between $\sigma_v$ and $\sigma_\mathrm{int}$ (i.e., increasing the former tends to lead to decreases in the latter).

\subsection{Residual Bulk-Flow Velocity}

Deferring a discussion of $\beta$ until the following section, we turn our focus to $\mathbf{V}_\mathrm{ext}$, the residual, coherent bulk-flow velocity arising from the gravitational interaction between the objects in our DSS set and large-scale structure existing beyond the volume encompassed by our reconstruction. Our result, that $V_\mathrm{ext} = 195_{-23}^{+22}$\,km\,s$^{-1}$ in the direction $(\ell, b) = (292_{-7}^{+7}, -6_{-4}^{+5})$\,deg, is listed in Table~\ref{tab:flow-model} along with the corresponding values for various subsets of our full DSS sample.

%
\begin{table}
{\addtolength{\tabcolsep}{-1.5pt}\renewcommand{\arraystretch}{1.25}
\caption[Flow model parameters]{Flow model parameters derived from the DSS and subsets of it.\label{tab:flow-model}}
\begin{tabular}{lrccrr}
\hline
\hline
Sample &  $N$ & $f\sigma_{8}^{\textrm{lin}}$ & $V_{\textrm{ext}}$ (km\,s$^{-1}$) &          $l$ (deg) &          $b$ (deg) \\
\hline
SNe Ia (WLR) &  540 &    $0.393_{-0.024}^{+0.024}$ &                $193_{-25}^{+25}$ &    $289_{-8}^{+8}$ &     $-1_{-6}^{+6}$ \\
SNe Ia (SDM)        &  137 &    $0.450_{-0.055}^{+0.057}$ &                $254_{-63}^{+67}$ &  $285_{-19}^{+18}$ &   $-2_{-11}^{+11}$ \\
SNe Ia (all)    &  677 &    $0.400_{-0.023}^{+0.023}$ &                $200_{-23}^{+24}$ &    $289_{-7}^{+7}$ &     $-2_{-5}^{+5}$ \\
SNe II     &   98 &    $0.388_{-0.063}^{+0.061}$ &                $184_{-59}^{+62}$ &  $319_{-29}^{+29}$ &  $-21_{-17}^{+16}$ \\
\hline
DSS        &  775 &    $0.390_{-0.022}^{+0.022}$ &                $195_{-23}^{+22}$ &    $292_{-7}^{+7}$ &     $-6_{-4}^{+5}$ \\
\hline
\end{tabular}}
\end{table}


Our result is mostly in agreement with \citet{A2}, who find\footnote{In the original journal publication, \citet{A2} quoted a significantly lower $V_\mathrm{ext}$ for the A2 sample, but this was the result of a systematic redshift error. A revision is underway with the journal that reflects the value we report herein.} $V_\mathrm{ext} = 177_{-26}^{+24}$\,km\,s$^{-1}$ toward $(\ell, b) = (289_{-8}^{+9}, 9_{-9}^{+9})$\,deg for the A2 compilation and similar (albeit more tightly constrained values) when they jointly fit A2 with a large sample of galaxy-derived peculiar velocities. Discrepancies between our result and theirs may stem from our use of a different dataset (both in the sense of using A2.1 instead of A2, and in the sense of including additional subcatalogues), slightly different priors (that we believe to be more valid for the domain of application), or our incorporation of SALT2-parameter uncertainties in Equations~\ref{eq:tripp-error} and~\ref{eq:tripp-SNII-error}. We note that our result is also consistent with (but again higher in magnitude than) that found by \citet{Carrick}. Our hope is that larger, homeogeneous samples of SNe~Ia and SNe~II may bring further convergence in the coming years.

\subsubsection{Bulk-Flow Velocity}

Distinct from the residual bulk-flow velocity, $\mathbf{V}_\mathrm{ext}$, discussed herein, many studies measure a ``bulk flow'' from their peculiar-velocity catalogues. The difference between the two is that the former is intended to be due \emph{solely} to structure beyond the 2M++ reconstruction, while the latter (which we refer to as $\mathbf{V}_\mathrm{bulk}$ for clarity) makes no such distinction. In order to facilitate a comparison between our work and the aforementioned studies, we derive an analogous bulk-flow measurement by repeating the procedures of Section~\ref{sec:method} except with the constraint that $\beta = 0$ enforced.

In effect, this ``turns off'' the peculiar-velocity reconstruction (as evidenced in Eqs.~\ref{eq:modified_perturbation}~\&~\ref{eq:zpred}), and with it being no longer able to contribute in the comparison to the observed peculiar velocities, all that remains is $\mathbf{V}_\mathrm{ext} \to \mathbf{V}_\mathrm{bulk}$. Following this approach, we find $V_\mathrm{bulk} = 245_{-31}^{+32}$\,km\,s$^{-1}$ in the direction $(\ell, b) = (294_{-7}^{+7}, 3_{-5}^{+6})$\,deg. 
%

\begin{figure}
    \includegraphics[width=\columnwidth]{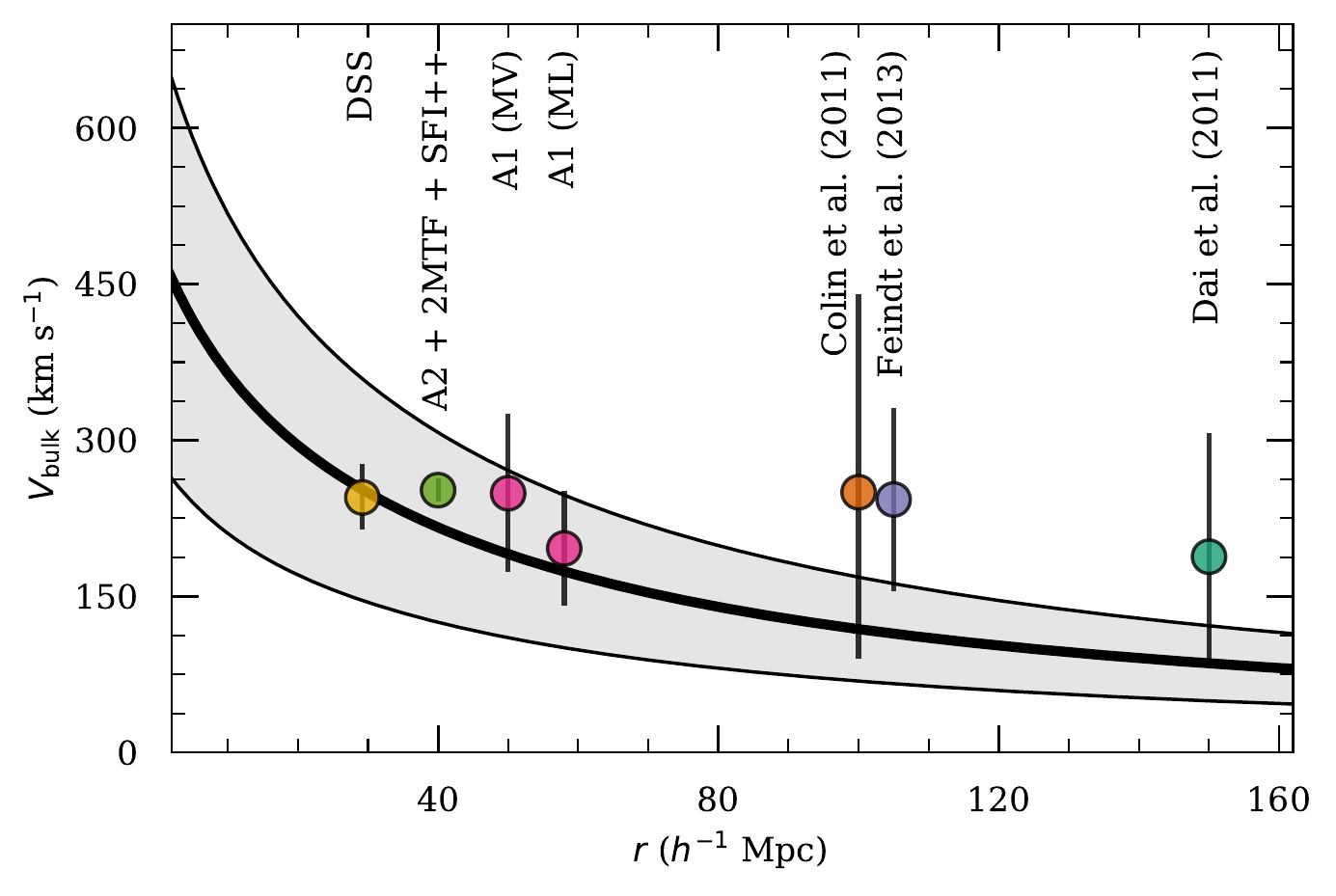}
    \caption{Our result for the bulk-flow amplitude (left-most point) compared with other SN-based results and the $\Lambda$CDM prediction (shown as the black line with 68\% confidence region). The two comparison values from the A1 compilation are derived via the \emph{minimum-variance} and \emph{maximum-likelihood} methods, respectively \citep[see][for more details]{A1}. Where results are quoted over a redshift shell instead of at a characteristic distance \citep[e.g.,][]{DaiBF,FeindtBF}, we convert the more distant end of the shell to the distance used in our comparison.}
    \label{fig:vbulkcompare}
\end{figure}

In comparing to other bulk-flow measurements (see Fig.~\ref{fig:vbulkcompare}), we caution that a variety of methods exist --- each with their own advantages and drawbacks \citep[see, e.g.,][for a discussion of two such methods; ours is of the \emph{maximum-likelihood estimate} variety]{A1}. Moreover (and independent of the method), each peculiar-velocity catalogue will have its own characteristic depth (e.g., ours is $\sim 30\,h^{-1}$\,Mpc), and comparisons should  only be made at comparable depths. Keeping these caveats in mind, we are contented to find consistency between our result and those of other SN-based studies, as well is with the $\Lambda$CDM expectation. Though we show only bulk-flow magnitudes in Figure~\ref{fig:vbulkcompare}, we find reasonable directional consistency as well.

\subsection{Constraint on $f\sigma_8$}
\label{ssec:fsig8}

Our result, $\beta = 0.418_{-0.020}^{+0.020}$, is best interpreted after transforming according to $f\sigma_8 = (f/b)(b\sigma_8) = \beta \sigma_8^g$. Here $\sigma_8^g = 0.99 \pm 0.04$ is the root-mean-squared fluctuation in the galaxy field as determined by \citet{Carrick}. We perform this operation directly on the $\beta$ posterior samples from our MCMC chains, resulting in $f \sigma_8 = 0.413_{-0.026}^{+0.026}$. 
We must, however, note that the value of $\sigma_8$ appearing in this equation is the measured, nonlinear value of the root-mean-square density fluctuation in an 8\,$h^{-1}$ Mpc sphere at $z \approx 0$. As a result, it cannot be directly compared with other measurements in the literature which refer to the linear value of this quantity (hereafter, $\sigma_{8}^{\mathrm{lin}}$) extrapolated to $z = 0$. Thus, we need to linearise our result prior to comparing it with constraints derived at high redshifts. To do so, we adopt the prescription outlined in Equation~3.13 of \citet{jlinearize},
\begin{equation}
\left(\sigma_{8}^{\mathrm{lin}}\right)^2 = \frac{\sqrt{1 + 0.864\sigma_8^2} - 1}{0.432} ,
\label{eq:linearise}
\end{equation}
which requires that we explicitly assume a value of $\Omega_m$ (we use 0.3, as stated in Sec.~\ref{sec:intro}) to break the degeneracy in $f\sigma_8$ to constrain $\sigma_8$ directly. We do this via $f \approx \Omega_m^{0.55}$, which, as introduced in Section~\ref{sec:intro}, is valid for $\Lambda$CDM cosmology, resulting in $\sigma_8 = 0.801_{-0.050}^{+0.050}$. After linearising according to Equation~\ref{eq:linearise}, this gives $\sigma_8^\mathrm{lin} = 0.756_{-0.043}^{+0.043}$. Similar results are obtained using the \citet{MeadBriedenTroster2021} nonlinear modifications to the power spectrum (without baryonic feedback) to calculate $\sigma_8$ and $\sigma_8^\mathrm{lin}$. Hence, we find $f\sigma_{8}^{\mathrm{lin}} = 0.390_{-0.022}^{+0.022}$, representing the tightest SN-based constraint on the quantity to date. As with $\mathbf{V}_\mathrm{ext}$, we summarise our DSS-derived result as well as those derived from subsets of it in Table~\ref{tab:flow-model}.

\begin{figure*}
    \includegraphics[width=\textwidth]{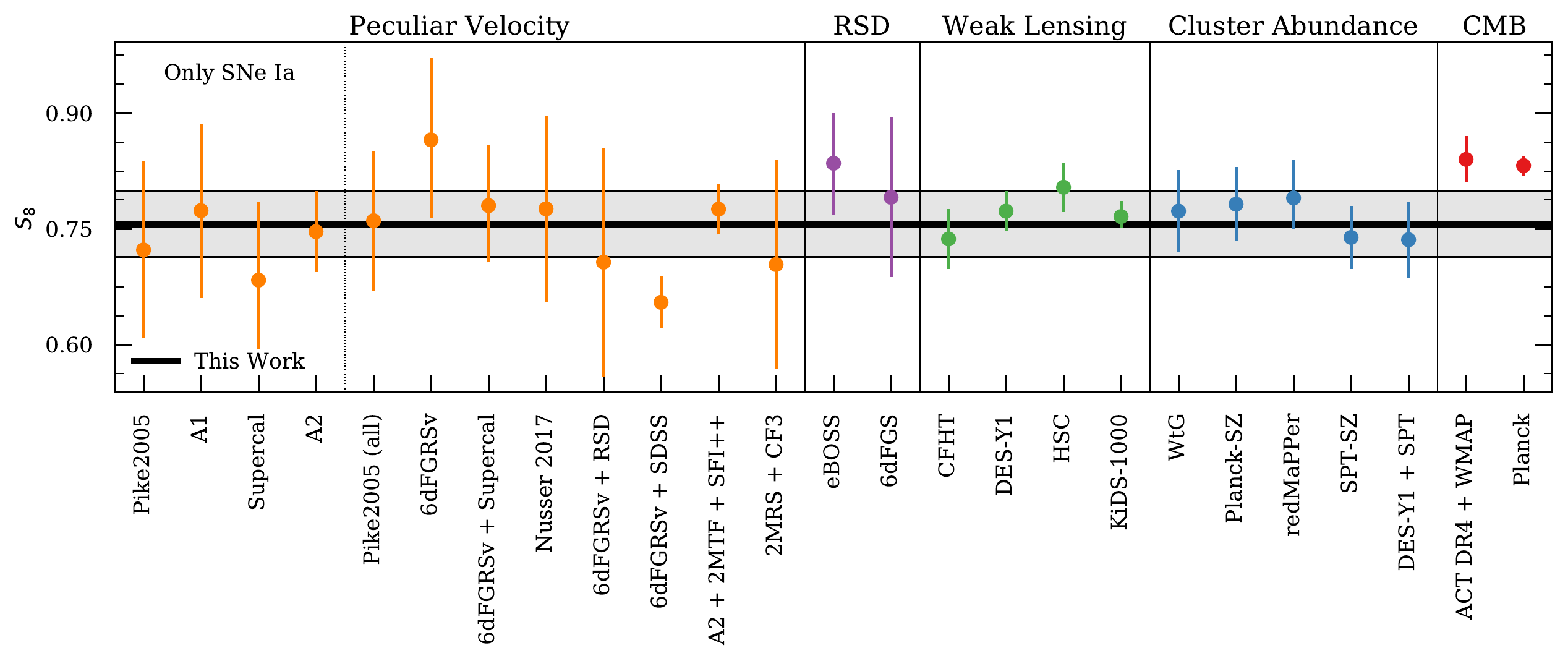}
    \caption{Our result for $S_8$ (horizontal line) compared with results from the literature derived via multiple means (as delineated at the top of each distinct section). All names and values used in the RSD, Weak Lensing, Cluster Abundance (except DES-Y1 + SPT), and CMB (except ACT DR4 + WMAP) sections are as listed in Table 7 of \citet{A2}, as are Nusser2017, 6dFGRSv + RSD, and 6dFGRSv + SDSS of the Peculiar Velocity section. All others are as discussed herein. References are as follows (in order of first appearance from left to right): \citet[Pike2005;][]{fwd_lkl}, \citet[A1;][]{A1}, \citet[Supercal and/or 6dFGRSv;][]{Huterer_SN}, \citet[A2;][]{A2}, \citet[Nusser2017;][]{Nusser2017}, \citet[6dFGRSv + RSD;][]{AB20}, \citet[6dFGRSv + SDSS;][]{Said_Gal}, \citet[2MRS + CF3;][]{2MRS+CF3}, \citet[eBOSS;][]{eBOSS}, \citet[6dFGS;][]{Beutler2012}, \citet[CFHT;][]{Heymans2013}, \citet[DES-Y1;][]{DES2018}, \citet[HSC;][]{HSC}, \citet[KiDS-1000;][]{Heymans2020}, \citet[WtG;][]{WtG}, \citet[Planck-SZ;][]{Planck-SZ}, \citet[redMaPPer;][]{redMaPPer}, \citet[SPT-SZ;][]{SPT-SZ}, \citet[DES-Y1 + SPT;][]{DESSPT}, \citet[ACT DR4 + WMAP;][]{ACT+WMAP}, and \citet[Planck;][]{Planck}.}
    \label{fig:S8}
\end{figure*}

Our result is consistent with both the A2-only and A2 + galaxy constraints derived by \citet{A2}, albeit closer to the latter. We find this encouraging --- it would seem that our larger sample of SNe (including SDM-treated SNe~Ia and SNe~II) is on the path to converging with comprehensive, large-scale studies utilising thousands of galaxies. Indeed, we find this to be the case after comparing to a vast array of studies that utilise multiple methods (see Fig.~\ref{fig:S8}).

Instead of $f\sigma_8$, however, we use a related quantity for this comparison: $S_8 \equiv \sigma_8(\Omega_m/0.3)^{0.5} \approx f\sigma_8^\mathrm{lin}/(0.3^{0.55})$, where the equivalency denotes the definition and the approximation denotes its mapping from our results. In doing so, we can easily compare our result to those derived through a variety of observations including the cosmic microwave background (CMB), cluster abundances, weak gravitational lensing, redshift-space distortions (RSDs), and of course peculiar velocities (as we have used in this work). In all but the last category (i.e., peculiar velocities) and the cases of DES-Y1 + SPT and ACT DR4 + WMAP \citep[which we take directly from the publications;][respectively]{DESSPT,ACT+WMAP}, we obtain $S_8$ values directly from Table 7 of \citet{A2}, who have already taken the requisite steps to convert and homogenise the values reported by the original studies. This remains true for the peculiar-velocity category as well, excepting the following.
\begin{enumerate}
    \item Pike2005: \citet{fwd_lkl} report $\sigma_8 (\Omega_m/0.3)^{0.6} = 0.86 \pm 0.15$ for their SN~Ia-only sample and $\sigma_8 (\Omega_m/0.3)^{0.6} = 0.91 \pm 0.12$ for their full compilation (including SNe~Ia and galaxies). We convert both to $S_8$ by means of a multiplicative factor of $(\Omega_m/0.3)^{-0.1}$ and the linearisation procedure described above, prior to including them in Figure~\ref{fig:S8}.
    \item A1: \citet{A1} derive $f\sigma_8 = 0.424 \pm 0.069$ from their A1 SN~Ia sample. We convert this to $S_8$ identically to our own result after linearisation. 
    \item Supercal and/or 6dFGRSv: \citet{Huterer_SN} produce three constraints on $f\sigma_8$: $0.370_{-0.053}^{+0.060}$ for the ``Supercal'' sample of SNe~Ia \citep{supercal}, $0.481_{-0.064}^{+0.067}$ for a large set of fundamental-plane distances from the 6dF galaxy survey \citep{6dF}, and $0.428_{-0.045}^{+0.048}$ for both samples combined. Our conversion of each to $S_8$ is identical to our treatment of A1 and our own result.
    \item A2: Although \citet{A2} provide $S_8$ for their full compilation (which we refer to in Fig.~\ref{fig:S8} as A2 + 2MTF + SFI++), they do not provide the conversion for just the A2 sample. Thus, we convert the relevant value in order to include it in Figure~\ref{fig:S8}.
    \item 2MRS + CF3: By comparing constrained realisations of the peculiar-velocity field from the Two-Micron All-Sky Redshift Survey \citep[2MRS;][]{2MRS-Huchra,2MRS-Macri} with observed peculiar velocities from the Cosmicflows-3 catalogue \citep[CF3;][]{CF3}, \citet{2MRS+CF3} derive $f\sigma_8^\mathrm{lin} = 0.363 \pm 0.070$. We convert this to $S_8$ identically to our own result.
\end{enumerate}
As Figure~\ref{fig:S8} shows, our result is highly consistent with those of other peculiar-velocity-based studies, along with those derived across multiple other methods (except for CMB-based measurements, which we discuss below). Indeed, even the most egregious disagreement \citep[6dFGRSv + SDSS;][]{Said_Gal} amounts to $< 1.9\sigma$ --- comfortably below the threshold for statistical significance --- and only two more out of a total of 25 exceed the $1\sigma$ discrepancy level, though we emphasise that the samples used in the comparison are not all mutually independent. Moreover, our result --- derived with $< 800$ SNe --- is more tightly constrained (based on its statistical-only error bars) than all but two other peculiar-velocity-based constraints, both of which use many thousands of galaxy distances (and in one case, the A2 SN~Ia sample as well). Thus, in the coming era of wide-field, relatively high-cadence surveys, the prospect for SNe (i.e., well-sampled SNe~Ia standardised with a WLR, and now SNe~II as well as sparsely observed SNe~Ia standardised with the SDM) to produce the best low-redshift constraint on $f\sigma_8$ is particularly bright.

\subsubsection{Aside: The $S_8$ Tension}

As an interesting aside, we note that the single most tightly constrained value in our comparison \citep[i.e., \textit{Planck};][]{Planck} is one of the two remaining values at $> 1\sigma$ tension with our result (and the other, ACT DR4 + WMAP, is also CMB-based). Given our result's consistency with those it is compared against, this means that the CMB results are in modest tension with \emph{most} other comparison values. In fact, the second most tightly constrained \citep[i.e., KiDS-1000;][]{Heymans2020} value we consider --- which is in full agreement (i.e., $0.2 \sigma$ difference) with our result --- differs from the \textit{Planck} value at the $\sim 3 \sigma$ level. Moreover, if we pull out the \textit{Planck} result and compare it with the inverse-variance-weighted mean of all other values (including ours), the tension reaches $4.1\sigma$ and remains $\gtrsim 3\sigma$ when we compare it to the same aggregation applied separately to the Peculiar Velocity ($4.5\sigma$), Weak Lensing ($3.3\sigma$), and Cluster Abundance ($2.8 \sigma$) categories.

Though these levels are overestimated in some cases (e.g., there is overlap in the datasets used to derive some of the comparison values, and there is no guarantee that the comparison values we have used are a truly comprehensive sample), we would be remiss not to note the parallels to the current H$_0$ tension \citep[e.g.,][]{RiessReview} --- once again, we have a cosmological parameter whose value as measured with low-redshift data is in significant disagreement with the value inferred from CMB measurements. The difference here is that, discounting the 6dFGRSv + SDSS result (which appears to be an outlier in the Peculiar Velocity category), it is only the \textit{aggregation} of multiple measurements \citep[with the marginal exception of KiDS-1000; see][for more discussion on this ``$S_8$ tension'']{S8tension} that is in tension, and thus we \emph{do not} escalate our findings in this area beyond simply noting them here. As with other parts of our analysis, we are optimistic that larger, homogeneous samples of SNe~Ia and II studied using our methodology will be able to shed more light on this issue.

\section{Conclusion}
\label{sec:conclusion}

In this paper, we painstakingly assemble the largest-ever SN-based peculiar-velocity catalogue: the Democratic Samples of Supernovae,  consisting of 775 objects. In addition to its sheer size, the DSS is novel for (i) its ``resurrection'' of otherwise unusable SN~Ia observations (owing to data sparsity) via the snapshot distance method, and (ii) its inclusion of SNe~II, which have not until now been used in such a way. Our SDM-enabled subcatalogue of 137 objects represents the final distillation of a candidate pool that started at $\sim 2,500$ objects; with future improvements to the SDM and the tools that enable it, the efficiency may well increase significantly. On this and both other fronts (SNe~II and conventional SN~Ia distances), significantly larger samples than our DSS should rapidly become feasible with the prevalence of wide-field surveys in the coming years that will discover hundreds of thousands of SNe. Our intent is for the work described herein to become the prototype for future studies that leverage such upcoming datasets.

To draw inferences from our DSS catalogue, we update and utilise a forward-likelihood framework that has been used in related works for the last $\sim 15$\,yr. In this approach, a parameter ($\beta = f/b$) related to $f\sigma_8$ and the magnitude and direction of a coherent, external bulk flow are jointly fit with subcatalogue-specific nuisance parameters that serve to cross-calibrate between distinct subcatalogues. After performing basic validity checks on the fitted nuisance parameter values (all of which are satisfactory), we report top-level results of $f\sigma_{8}^{\mathrm{lin}} = 0.390_{-0.022}^{+0.022}$ and $V_\mathrm{ext} = 195_{-23}^{+22}$\,km\,s$^{-1}$ in the direction $(\ell, b) = (292_{-7}^{+7}, -6_{-4}^{+5})$\,deg --- the tightest SN-based constraints (considering statistical error bars only) ever produced on each. Moreover, we find a bulk flow (ignoring any influence from the peculiar-velocity reconstruction) of $V_\mathrm{bulk} = 245_{-31}^{+32}$\,km\,s$^{-1}$ toward $(\ell, b) = (294_{-7}^{+7}, 3_{-5}^{+6})$\,deg on a scale of $\sim 30\,h^{-1}$\,Mpc.

By converting $f\sigma_8$ to $S_8$, we demonstrate that our result is consistent with that of many other studies leveraging multiple methodologies. Indeed, none deviate from ours at a level that warrants statistical significance, though we do find the \textit{Planck} value \citep{Planck} --- which differs from our result by $1.7\sigma$ --- to be significantly different from aggregations of our value with other comparisons. Our result for $\mathbf{V}_\mathrm{ext}$ is consistent in direction with prior studies \citep[e.g.,][]{A2}, but falls at the high end of magnitudes in the same studies (though still consistent with the central values). We discuss this briefly, but ultimately conclude that a larger, more homogeneous SN sample will be instrumental in clarifying the matter. Consistent with what is stated above, we believe that such samples will become available in the near future, offering an answer to this question while perhaps opening others. Indeed, the future of SN~Ia and II cosmology, particularly in the era of wide-field surveys, holds much promise.

\section*{Acknowledgements}

We thank our referee, Dragan Huterer, for constructive feedback that meaningfully contributed to the quality of this work. B.E.S. thanks K. D. Zhang for illuminating discussions on Bayesian statistics and Marc J. Staley for providing fellowship funding.
A.V.F. has been generously supported by the TABASGO Foundation, the Christopher R. Redlich Fund, Gary \& Cynthia Bengier (T.d.J. was a Bengier Postdoctoral Fellow at U.C. Berkeley), and the Miller Institute for Basic Research in Science (in which he is a Senior Miller Fellow).
This research used the Savio computational cluster resource provided by the Berkeley Research Computing program at U.C. Berkeley (supported by the Chancellor, Vice Chancellor for Research, and Chief Information Officer).

\section*{Data Availability}

The Democratic Samples of Supernovae (DSS), which mostly comprises previously published data (as documented and referenced in Sec.~\ref{sec:data}), will be provided by the authors upon request. Distances derived from the DSS are also available in the Extragalactic Distance Database\footnote{\url{https://edd.ifa.hawaii.edu/}} \citep{EDD}.




\bibliographystyle{mnras}
\bibliography{references} 




\appendix

\section{Snapshot Distance Sample Selection}
\label{app:sdm}

Beyond the cuts employed in selecting our initial SDM candidate pool (see Sec.~\ref{sssec:sdm}), many more are required to ensure sufficient quality and homogeneity, and thus we carefully document the sequence of cuts leading to our final sample below (the cuts are also summarised in Fig.~\ref{fig:cuts}). Given the sheer size of our initial sample (2900 SNe with $> 8900$ spectra and $>$ 136,000 photometric points), we judiciously select the order so that algorithmic cuts are always applied before those that require human attention.
\begin{description}
\item \textbf{Overlap:} Our first cut is to remove the nearly 400 SNe in our sample that also occur in A2.1 $+$ LOSS2.0. The fact that almost all SNe from the latter are present in the former is expected, and those that are not usually fail the requirement that at least one optical spectrum be available on the OSC. After imposing this cut, 2507 SNe remain with $\sim 4700$ spectra and $\sim$ 70,000 photometric points. It is encouraging that the percentages of spectra and photometric points removed are both larger than the corresponding percentage of SNe --- those SNe that are removed are, on the average, much better sampled than those that remain, rendering support to our assertion that our A2.1 + LOSS2.0 amalgamation accounts for the majority of publicly-available SN~Ia observations.
\item \textbf{Recency:} To ensure that the data used herein were collected with relatively modern techniques and equipment (e.g., CCDs), we drop spectra and photometric points that were observed prior to 1990 (formally, we required that the MJD of all observations is $\geq 47892$). Since the vast majority of human-studied SNe have been discovered in recent years, this removes only a very modest number of objects, spectra, and photometric points.
\item \textbf{Spectral Coverage and Resolution:} As \texttt{deepSIP} forms a central component of the SDM, we must cut spectra that are not suitable for it. Specifically, this means that we must drop spectra that do not have full coverage\footnote{Following S20, we define ``full coverage'' as having a minimum wavelength of less than 5750\,\AA~and a maximum that exceeds 6600\,\AA.} of the \ion{Si}{ii} $\uplambda 6355$ feature, along with those having insufficient resolution. Though the former removes $\sim 300$ spectra, only 13 SNe are removed as a result. The latter is much more impactful, however, removing $\sim 700$ spectra \citep[mostly from the SED Machine;][]{SEDM} and nearly as many SNe. In fact, this amounts to the most severe cut on SNe and the second most severe cut on spectra (after the ``overlap'' cut). With future improvements to \texttt{deepSIP} it may be possible to revisit and lessen the extremity of this cut, but this will have to wait until those improvements can be realised.
\item \textbf{Irrelevant Photometric Passbands:} After the cuts described above, the 60,140 photometric points that remain are in one of 62 different passbands, some of which are irrelevant to our analysis and must therefore be removed. By far, the primary factor in rendering observations in a given passband ``irrelevant'' is if that band is too broad. For example, 4923 \textit{Gaia} $G$-band \citep{GaiaGband}, 2104 Pan-STARRS $w$-band \citep{PS1}, 612 ATLAS $c$-band or $o$-band \citep{ATLAS}, and many other wide-band observations are removed, along with a further 5061 unfiltered photometric points. A second factor in determining the relevance of a passband is the wavelengths of light that it passes. As our SDM uses a WLR implementation that is trained at optical and near-infrared wavelengths, we must discard passbands that lie outside this domain, thereby removing some \textit{Swift} observations \citep{Swift} amongst others. All told, 15,624 photometric points are removed, and those that remain are homogenised down to 21 distinct passbands.
\item \textbf{Photometric Uncertainties:} Of the remaining 44,516 photometric points, 10,429 do not include a symmetric uncertainty in the reported magnitude. We are able to ``fix'' 6120 of these by taking the larger of their (asymmetric) lower and upper magnitude uncertainty estimates, but 4393 remain which have no identifiable quantitative uncertainty indicator. We take the conservative approach of removing these data, leaving us with just under 1200 SNe covered by $\sim 2900$ spectra and $\sim$ 40,000 photometric points.
\item \textbf{$2+$ Distinct Passbands:} As the SDM requires, at a minimum, two photometric points in \emph{different} passbands (and a spectrum), we remove SNe that fail this requirement. This reduces the sample to 747 SNe with $\sim 2400$ spectra and $\sim$ 35,000 photometric points.
\item \textbf{Literature Search for Photometric Systems:} While we have taken care above to homogenise the passbands in our sample (in the sense of resolving the differing names given to identical filters), we must also understand the photometric systems to which they belong. To do so, we visit the primary source provided by the OSC for each remaining point in our photometry sample (amounting to 248 sources) and attempt to determine the photometric system. We find that 26 SNe (covered by 5719 photometric points) are either not an SN~Ia or are not an SN~Ia subclass that can be handled by our WLR. These are dropped, as are an additional 7220 photometric points whose photometric system is either not clear or not supported by our light-curve fitter \citep[consistent with LOSS2.0 distances, we use the ``EBV\_model'' in \texttt{SNooPy};][]{SNooPy}. This leaves 21,743 photometric points remaining (covering 670 SNe) in 16 distinct photometric systems, with the plurality being the standard Landolt system \citep{Landolt1983,Landolt1992}, followed by the CSP natural system \citep{CSP1}, and then by the \textit{Swift} photometric system \citep{Swift}. To maximise the amount of data we can use, we have registered the photometric systems of the Zwicky Transient Facility \citep{ZTF} and LOSS 1.0 \citep{Ganeshalingam2010} into our light-curve fitter.
\item \textbf{$2+$ Distinct Passbands:} As the cut described above modifies the photometric coverage for some objects in our sample, we must repeat the $2+$ passband cut. Though this repetition is, perhaps, aesthetically unappealing, it represents a significant savings of human time compared to the option of performing it only once at this stage (we would have needed to review many additional sources in the literature-search stage). Upon repeating this cut, we are left with 625 SNe covered by $\sim 1900$ spectra and $\sim$ 21,000 photometric points.
\item \textbf{Redshifts:} As the purpose of this paper is to study the peculiar velocity field, precise redshifts are of paramount importance \citep{DavisRedshiftPV}. We therefore devote significant effort to obtaining high-quality host-galaxy redshifts for the SNe in our sample using the following strategy. First, we query NED\footnote{The NASA/IPAC Extragalactic Database (NED) is operated by the Jet Propulsion Laboratory, California Institute of Technology, under contract with the National Aeronautics and Space Administration.} for the (heliocentric) redshift of the host galaxy of each SN in our sample. For those that fail (either due to the host galaxy not having a listed redshift, or the host galaxy not being provided by the OSC), we perform a careful literature search, accepting only host-galaxy-derived (and not SN-feature-derived) redshifts. This reduces our sample to 480 objects with reliable, high-quality redshifts. We convert each (heliocentric) redshift to the CMB frame and then (consistent with A2.1 + LOSS2.0), for those objects known to be in a cluster of galaxies, we update the redshift to that of the cluster.
\item \textbf{deepSIP:} At this stage, we deploy \texttt{deepSIP} on the spectra that remain in our sample to derive the phase and light-curve shape measurements required by the SDM. Of the 1752 spectra processed by \texttt{deepSIP}, 631 are categorised as ``in-domain'' --- a requirement of the SDM --- leaving us 243 SNe with which to proceed with our analysis. For those SNe having multiple spectra, we assign their time of maximum brightness and light-curve shape as the average of the relevant \texttt{deepSIP}-reconstructed quantities. As with other cuts that stem from \texttt{deepSIP}, the severity of this cut can likely be reduced in the future as the model is improved.
\item \textbf{Photometric Phase:} Using the \texttt{deepSIP}-inferred times of maximum light, we compute the rest-frame phase of every photometric point in our dataset, and remove any that are earlier than $-10$\,d or later than $70$\,d, corresponding to the temporal domain over which \texttt{SNooPy} models light curves. This reduces our sample only modestly, leaving 11,063 photometric points covering 232 SNe.
\item \textbf{$2+$ Distinct Passbands:} As the cut described above modifies the photometric coverage for some objects in our sample, we must again repeat the $2+$ passband cut. This reduces our sample by a further nine objects, leaving 223 SNe covered by 610 spectra and 11,034 photometric points.
\end{description}
Those objects that remain at this point are then processed in accordance with the description in Section~\ref{sssec:sdm}.


\bsp	
\label{lastpage}
\end{document}